# Comments on identity of gap anisotropy and nodal constant Fermi velocity in $Bi_2Sr_2CaCu_2O_{8+\delta}$

Hyun-tak Kim (htkim@etri.re.kr), Bong-jun Kim, Kwang-yong Kang
Convergence & Components Lab., MIT Device Team, ETRI, Daejeon 305-350, Korea

On the basis of both the inhomogeneity of a superconductor $Bi_2Sr_2CaCu_2O_{8+\delta}$ and the effect of measurement, we reveal that an anomalously large gap anisotropy known as evidence of $d_{x^2-y^2}$-wave symmetry [Phys. Rev. Lett. 70 (1993) 1553] is not intrinsic, and that the constant Fermi velocity at node as an unsolved problem [Nature 423 (2003) 398] is due to the $d_{x^2-y^2}$-wave insulator-metal transition.

An anomalously large gap anisotropy measured in the a-b plane of $Bi_2Sr_2CaCu_2O_{8+\delta}$ (Bi-2212) has been reported as evidence of $d_{x^2-y^2}$-wave symmetry [1]. Many researchers have thus far accepted this explanation. However, the superconducting gap in the tunneling measurement had the s-wave symmetry with the ?-type near zero voltage [2]. Moreover, the invarance of the Fermi velocity at the node in cuprates remains unsolved [3,4]. Therefore, the pairing symmetry and the constant Fermi velocity need to be newly investigated for an inhomogeneous system.

It has been known that Bi-2212 is intrinsically inhomogeneous, and that the homogeneous region is within 3nm [5]. Because the diameter of the X-ray beam for angle-resolved photoemission spectroscopy (ARPES) does not reach 3nm, experimental data include the effects of two phases of metal (region B; superconductor below $T_c$) and an insulator (region A) (Fig. 1a). If only region A or B is measured, a pseudogap of 300 meV [6] or an exact superconducting gap will be observed, respectively; no proximity effect is assumed [7]. In real measurement, a local carrier density (band filling, the extent of the metal phase in Fig. 1a) becomes $0<\rho=n/m<1$ where $n$ is the number of carriers and $m$ is the number of atoms (right of Fig. 1a), because the metal region (left of Fig. 1a) is averaged over the measurement region. An observed energy gap, $\Delta_{obs}$, is given by

$$\Delta_{obs} = \Delta_{intrinsic}/\rho, \qquad (1)$$

where $\Delta_{intrinsic}$ is an intrinsic superconducting true gap and $0<\rho\leq 1$ [8,9]. $\Delta_{obs}$ increases with decreasing $\rho$ and is an average value of $\Delta_{intrinsic}$ over the measurement region; the physical meaning of $\Delta_{obs}$ is the effect of measurement. $\Delta_{intrinsic}$ is constant irrespective of the extent of $\rho$. An application of Eq. 1 is shown in Fig. 1b. When $\rho\neq 1$, anisotropy of $\Delta_{intrinsic}$ cannot be observed.

Figure 1c shows a shift of the superconducting gap to the Fermi level at a node in curve B, as indicated by arrows, although the gap peak at the node is not clear because of the low energy resolution of ARPES [1]. The shift creates a difference in the energy-gap magnitude between the node and the non-node, which is the gap anisotropy and has been cited as evidence of the $d$-wave symmetry [1]. It is suggested here that the pairing symmetry of the pseudogap of the hump in curve A at the non-node is $d_{x^2-y^2}$-wave, because the hump at the node disappears (dot line in Fig. 1c); this phase is regarded as the insulating $d$-wave phase shown in Fig. 1a. Note that inhomogeneity in real-space is easily determined when ARPES data with two gaps and over are transformed into real-space by a Fourier transform.

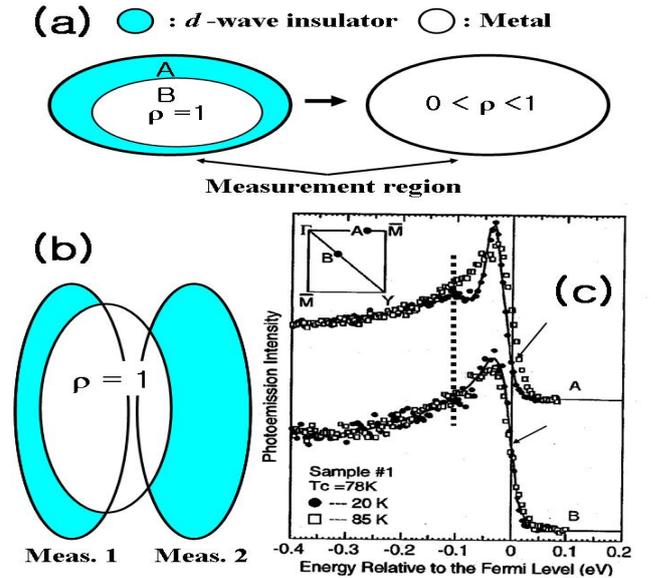

**Fig. 1a** Left: An inhomogeneous high-$T_c$ superconductor is composed of a metal phase (superconductor below $T_c$) and a $d$-wave insulator phase in real space. Right: When measured, the left system becomes right. **b**, Since Meas. 1 has a larger metal region than Meas. 2, $\rho_{meas.1}>\rho_{meas.2}$ and $\Delta_{obs,\,meas.1}<\Delta_{obs,\,meas.2}$ are given by Eq. 1 and Fig. 1a. This indicates that $\Delta_{obs}$ changes with the extent of the metal phase included in the measurement region although the same metal phase is measured. **c**, Data is cited from Shen *et al.*, [1]. This shows the absence of a hump at the node. The superconducting gap at the node is not zero.

Meanwhile, the superconducting gap remains small in curve B. Since bound charges in the *d*-wave insulator (region A) change to carriers at the node, the number of total carriers (carriers in the metal phase + carriers produced from the *d*-wave insulator) at the node becomes larger than the number of carriers in the metal phase at the non-node; the $d_{x^2-y^2}$-wave insulator-metal transition occurs at the node and $\rho_{node} > \rho_{non-node}$. Then, since $\rho_{node} \approx 1$ is evaluated, the Fermi velocity for observation ($V_{F,obs} = \rho^{1/3} V_{F,intrinsic}$) [10] becomes constant; this can be interpreted as the invariant nodal Fermi velocity which was suggested by Zhou *et al.*[3] and has been unclarified up to now.

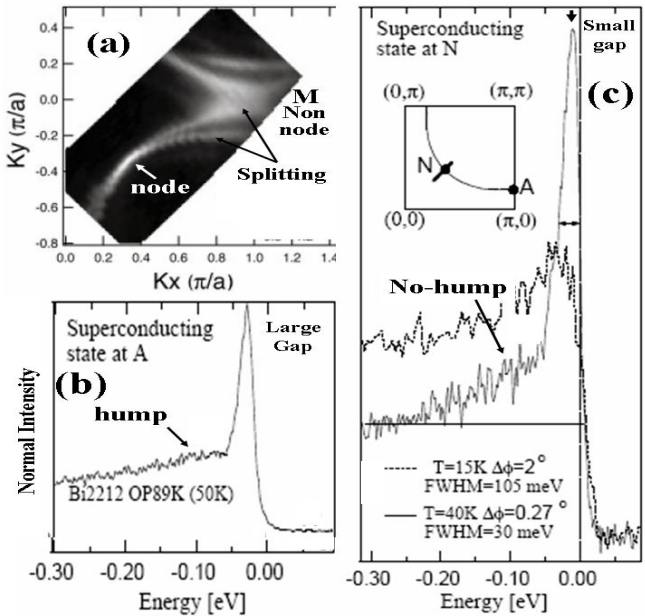

**Fig. 2 (a)** This was measured at 20 K (superconducting state) by high resolution ARPES for Pb-doped Bi-2212 [11]. The band splitting was explained by Bogdanov *et al.* [11]. Fig. (b) and (c) were measured by ARPES for Bi-2212 and cited from Kaminski *et al.* [12]. The large gap at the non-node (Fig. b) and the small energy gap at the node without hump (Fig. c) is shown.

As experimental evidence, it was observed that, although the band splitting at the non-node of M site is shown (Fig. 2), the spectral intensity at the node is thinner and brighter than that at the non-node (Fig 2a); this indicates $\rho_{node} > \rho_{non-node}$. This interpretation was not given by Bogdanov *et al.* [11]. The small gap peak without hump at the node was also measured [12,13] (Fig. 2c). When $\rho_{node} > \rho_{non-node}$ is applied to Eq. (1), $\Delta_{obs, node} < \Delta_{obs, non-node}$ is satisfied. The small superconducting gap is close to the intrinsic true superconducting gap. The gap anisotropy is attributed to the inhomogeneity of the metal phase (s-wave superconductor at low temperatures) and the *d*-wave insulator phase in the measurement region.

Thus, the observed gap anisotropy in reference 1 is not the anisotropy of $\Delta_{intrinsic}$ but rather an effect of measurement; this is not intrinsic. The constant Fermi velocity at the node in the normal state is due to the $d_{x^2-y^2}$-wave insulator-metal transition.

We thank Prof. A. Kaminski for valuable comments and permission to use experimental data in Fig. 2 (b) and (c).

**References**
[1] Z. X. Shen, D. S. Dessau, D. M. King, W. E. Spicer, A. J. Arko, D. Marshall, L. W. Lombardo, A. Kapitulnik, P. Dickinson, S. Doniach, J. DiCarlo, A. G. Loeser, C. H. Park, Phys. Rev. Lett. 70 (1993) 1553.
[2] K. Kitazawa, T. Hasegawa, and H. Sugawara, J. Kor. Phys. Soc. 31 (1997) 27.
[3] X. J. Zhou, T. Yoshida, A. Lanzara, P. V. Bogdanov, S. A. Kellar, K. M. Shen, W. L. Yang, F. Ronning, T. Sasagawa, T. Kakeshita, T. Noda, H. Eisaki, S. Uchida, C. T. Lin, F. Zhou, J. W. Xiong, W. X. Ti, Z. X. Zhao, A. Fujimori, Z. Hussain, Z. X. Shen, Nature 423 (2003) 398.
[4] Y. S. Lee, K. Segawa, Z. Q. Li, W. J. Padilla, M. Dumm, S. V. Dordevic, C. C. Homes, Y. Ando, and D. N. Basov, Phys. Rev. B72 (2005) 54529.
[5] K. M. Lang, V. Madhavan, J. E. Hoffman, E. W. Hudson, H. Eisaki, S. Uchida, and J. C. Davis, Nature 415 (2002) 412.
[6] F. Ronning, C. Kim, D. L. Feng, D. S. Marshall, A. G. Loeser, L. L. Miller, J. N. Eckstein, I. Bozovic, Z. X. Shen, Science 282 (1998) 2067.
[7] I. Bozovic, G. Logvenov, M. A. J. Verhoeven, P. Caputo, E. Goldobin, and T. H. Geballe, Nature 422 (2003) 873.
[8] H. T. Kim, J. Phys. Soc. Jpn. 71 (2002) 2106.
[9] H. T. Kim, Physica C399 (2003) 48.
[10] When the carrier density in the Fermi-velocity formula is replaced by $\rho m$, $V_{F,obs}$ for observation is obtained. m is the number of total atoms.
[11] P. V. Bogdanov, A. Lanzara, X. J. Zhou, S. A. Kellar, D. L. Feng, E. D. Lu, H. Eisaki, J. I. Shimoyama, K. Kishio, H. Hussain, and Z. X. Shen, Phys. Rev. B64 (2001) 180505.
[12] A. Kaminski, J. Mesot, H. Fretwell, J. C. Campuzano, M. R. Norman, M. Randeria, H. Ding, T. Sato, T. Takahashi, T. Mochiku, K. Kadowaki, and H. Hoechst, Phys. Rev. Lett. 84 (2000) 1788.
[13] T. Valla, A. V. Fedorov, P. D. Johnson, B. O. Wells, S. L. Hulbert, Q. Li, G. D. Gu, and N. Koshizuka, Science 285 (1999) 2110.